\documentclass[aps,prl,twocolumn,superscriptaddress,reprint,nofootinbib]{revtex4-2}

\usepackage{graphicx}% Include figure files
\usepackage{dcolumn}% Align table columns on decimal point
\usepackage{bm}% bold math
\usepackage[dvipsnames]{xcolor}
\usepackage{siunitx}
\usepackage[hidelinks]{hyperref}% add hypertext capabilities
\usepackage{xcolor}
\usepackage{amsmath}
\usepackage{float}
\usepackage{cleveref}
\usepackage[version=3]{mhchem}
\usepackage{booktabs}
\usepackage{multirow}
\usepackage{tikz}
\usetikzlibrary{decorations.pathreplacing}
\usetikzlibrary{positioning}
\usetikzlibrary{svg.path}

\definecolor{myred}{RGB}{220,74,55}

\newcommand{\nonpolar}[1][]{%
\color{myred}
\raisebox{-0.25ex}
{\begin{tikzpicture}[#1]%
  \begin{scope}
    \clip (-1.4ex,-0.4ex) rectangle (1.6ex,2.0ex);
    \draw rectangle(1.0ex,1.0ex);
    \draw (0.5ex, -0.4ex) -- (0.5ex, 1.4ex);
    \draw (-0.4ex, 0.5ex) -- (1.4ex, 0.5ex);
  \end{scope}
\end{tikzpicture}}
\color{black}%
}

\newcommand{\cis}{\emph{cis}}
\newcommand{\trans}{\emph{trans}}
\newcommand{\fac}{\emph{fac}}
\newcommand{\mer}{\emph{mer}}
\newcommand{\eps}{\varepsilon_\mathrm{r}}

\DeclareSIUnit \electron{\text{e}}
\DeclareSIUnit \angstrom{\text{\AA}}
\DeclareSIUnit \atom{\text{atom}}

\begin{document}

\sisetup{per-mode=power, bracket-unit-denominator=true, sticky-per}%

% \preprint{APS/123-QED}

\title{Anion-polarisation--directed short-range-order in antiperovskite \ce{Li2FeSO}}

\author{Samuel W. Coles}
\email{swc57@bath.ac.uk}
\affiliation{Department of Chemistry, University of Bath, Claverton Down, BA2 7AY, United Kingdom}
\affiliation{The Faraday Institution, Quad One, Harwell Science and Innovation Campus, Didcot, OX11 0RA, United Kingdom}

\author{Viktoria Falkowski}
\affiliation{Department of Chemistry, University of Oxford, Inorganic Chemistry Laboratory, Oxford OX1 3QR, United Kingdom}
\affiliation{The Faraday Institution, Quad One, Harwell Science and Innovation Campus, Didcot, OX11 0RA, United Kingdom}

\author{Harry S. Geddes}
\affiliation{Department of Chemistry, University of Oxford, Inorganic Chemistry Laboratory, Oxford OX1 3QR, United Kingdom}
\affiliation{The Faraday Institution, Quad One, Harwell Science and Innovation Campus, Didcot, OX11 0RA, United Kingdom}

\author{Gabriel E. P\'erez}
\affiliation{ISIS Neutron and Muon Source, STFC Rutherford Appleton Laboratory, Didcot OX11 0QX, United Kingdom}
\affiliation{The Faraday Institution, Quad One, Harwell Science and Innovation Campus, Didcot, OX11 0RA, United Kingdom}

\author{Samuel G. Booth}
\affiliation{Department of Materials Science and Engineering, University of Sheffield, Sheffield S1 3JD, United Kingdom}
\affiliation{The Faraday Institution, Quad One, Harwell Science and Innovation Campus, Didcot, OX11 0RA, United Kingdom}

\author{Alexander G. Squires}
\affiliation{Department of Chemistry, University of Bath, Claverton Down, BA2 7AY, United Kingdom}
\affiliation{Department of Chemistry, University College London, London WC1H 0AJ, United Kingdom}
\affiliation{The Faraday Institution, Quad One, Harwell Science and Innovation Campus, Didcot, OX11 0RA, United Kingdom}

\author{Conn O'Rourke}
\affiliation{Department of Chemistry, University of Bath, Claverton Down, BA2 7AY, United Kingdom}
\affiliation{The Faraday Institution, Quad One, Harwell Science and Innovation Campus, Didcot, OX11 0RA, United Kingdom}

\author{Kit McColl}
\affiliation{Department of Chemistry, University of Bath, Claverton Down, BA2 7AY, United Kingdom}
\affiliation{The Faraday Institution, Quad One, Harwell Science and Innovation Campus, Didcot, OX11 0RA, United Kingdom}

\author{Andrew L. Goodwin}
\affiliation{Department of Chemistry, University of Oxford, Inorganic Chemistry Laboratory, Oxford OX1 3QR, United Kingdom}
\affiliation{The Faraday Institution, Quad One, Harwell Science and Innovation Campus, Didcot, OX11 0RA, United Kingdom}

\author{Serena A. Cussen}
\affiliation{Department of Materials Science and Engineering, University of Sheffield, Sheffield S1 3JD, United Kingdom}
\affiliation{The Faraday Institution, Quad One, Harwell Science and Innovation Campus, Didcot, OX11 0RA, United Kingdom}

\author{Simon J. Clarke}
\affiliation{Department of Chemistry, University of Oxford, Inorganic Chemistry Laboratory, Oxford OX1 3QR, United Kingdom}
\affiliation{The Faraday Institution, Quad One, Harwell Science and Innovation Campus, Didcot, OX11 0RA, United Kingdom}

\author{M. Saiful Islam}
\affiliation{Department of Chemistry, University of Bath, Claverton Down, BA2 7AY, United Kingdom}
\affiliation{Department of Materials, University of Oxford, Oxford, OX1 3PH, United Kingdom}
\affiliation{The Faraday Institution, Quad One, Harwell Science and Innovation Campus, Didcot, OX11 0RA, United Kingdom}

\author{Benjamin J. Morgan}
\email{b.j.morgan@bath.ac.uk}
\affiliation{Department of Chemistry, University of Bath, Claverton Down, BA2 7AY, United Kingdom}
\affiliation{The Faraday Institution, Quad One, Harwell Science and Innovation Campus, Didcot, OX11 0RA, United Kingdom}

% \collaboration{MUSO Collaboration}%\noaffiliation

\date{\today}% It is always \today, today,

\begin{abstract}
% \emph{10th Anniversary Statement:}
% We wish to congratulate the \emph{Journal of Materials Chemistry A} on its 10 year anniversary.
% The continued development of novel energy materials combined with an increased understanding of their key chemical properties are critical elements in the development of a green and sustainable energy future.
% For the past 10 years, the \emph{Journal of Materials Chemistry A} has provided a home for papers reporting diverse aspects of energy materials research.
% These papers have introduced new materials, have changed our understanding of existing classes of materials, and have steered how we think and enquire as scientific researchers. We congratulate the editorial team of their achievement, and look forward to a further 10 years of both reading and contributing to the \emph{Journal of Materials Chemistry A}.\\ 

% \emph{Abstract:}
Short-range ordering in cation-disordered cathodes can have a significant effect on their electrochemical properties.
Here, we characterise the cation short-range order in the antiperovskite cathode material \ce{Li2FeSO}, using density functional theory, Monte Carlo simulations, and synchrotron X-ray pair-distribution-function data.
We predict partial short-range cation-ordering, characterised by favourable \ce{OLi4Fe2} oxygen coordination with a preference for polar \cis-\ce{OLi4Fe2} over non-polar \trans-\ce{OLi4Fe2} configurations.
This preference for polar cation configurations produces long-range disorder, in agreement with experimental data.
The predicted short-range-order preference contrasts with that for a simple point-charge model, which instead predicts preferential \trans-\ce{OLi4Fe2} oxygen coordination and corresponding long-range crystallographic order.
The absence of long-range order in \ce{Li2FeSO} can therefore be attributed to the relative stability of \cis-\ce{OLi4Fe2} and other non-\ce{OLi4Fe2} oxygen-coordination motifs.
We show that this effect is associated with the polarisation of oxide and sulfide anions in polar coordination environments, which stabilises these polar short-range cation orderings.
We propose similar anion-polarisation--directed short-range-ordering may be present in other heterocationic materials that contain cations with different formal charges.
Our analysis also illustrates the limitations of using simple point-charge models to predict the structure of cation-disordered materials, where other factors, such as anion polarisation, may play a critical role in directing both short- and long-range structural correlations.
\end{abstract}

\maketitle

%\tableofcontents

\section{Introduction}

Crystallographically disordered materials, in which two or more heteroatomic species are distributed over otherwise equivalent sites, find use in several applications \cite{KageyamaEtAl_NatCommun2018, HaradaEtAl_AdvMater2019}, including solar cells \cite{ScanlonAndWalsh_ApplPhysLett2012, VealEtAl_AdvEnergyMater2015, ChenEtAl_EnergyEnvironSci2021,YongEtAl_JAmChemSoc2018, MangelisEtAl_PhysChemChemPhys2019}, hydrogen storage \cite{HirscherEtAl_JAlloysComp2020, MarquesEtAl_EnergyEnvironSci2021}, and lithium-ion batteries \cite{LeeEtAl_Science2014, ClementEtAl_EnergyEnvironSci2020, MinafraEtAl_InorgChem2020, Morgan_ChemMater2021,McCollEtAl_NatCommun2022}.
The material properties of crystallographically disordered systems depend not only on their stoichiometry and crystal structure, but also on the short-range configuration of their disordered heteroatoms \cite{KageyamaEtAl_NatCommun2018, CharlesEtAl_ChemMater2018, HaradaEtAl_AdvMater2019, JiEtAl_NatureComm2019}.
One such example is provided by cation-disordered lithium-ion cathode materials, in which the short-range cation configuration controls lithium-transport rates, charge and discharge behaviour, and redox properties \cite{LeeEtAl_Science2014, SeoEtAl_NatureChem2016, JiEtAl_NatureComm2019, CaiEtAl_Matter2021,McCollEtAl_NatCommun2022}.

A full understanding of heteroatomic materials requires both an accurate description of their short-range structures and an understanding of the physical principles that promote or inhibit specific short-range orderings.
Such mechanistic understanding is particularly valuable for technologically relevant materials, where targeted synthesis protocols that promote or inhibit particular local structure motifs may allow the optimisation of key material properties.

While many anion-disordered heteroanionic materials have been structurally well-characterised \cite{MorelockEtAl_ChemMater2013,HaradaEtAl_AdvMater2019, KageyamaEtAl_NatCommun2018,PilaniaEtAl_npjComputMater2020, JohnstonEtAl_ChemCommun2018,WolffAndDronskowski_JComputChem2008}, cation-disordered heterocationic materials have been generally less studied.
For heteroanionic materials, various general design rules have been proposed to explain particular examples of partial or full anion-ordering, based on electronic, strain, or electrostatic effects \cite{KageyamaEtAl_NatCommun2018,CharlesEtAl_ChemMater2018,HaradaEtAl_AdvMater2019,PilaniaEtAl_npjComputMater2020}.
For heterocationic materials, however, the factors that direct short-range order preferences are less well understood \cite{UrbanEtAl_PhysRevLett2017}.
% Structural studies of cation-disordered materials are therefore interesting not only to gain a better understanding of specific materials, but also because they might reveal general chemical principles that explain structural preferences across a range of materials.

% Understanding the factors that direct local structure in crystallographically disordered materials can also be helpful when trying to construct structural models with explicit site occupancies.
% Computational models are often used to predict or investigate material properties of interest, but these typically require as input structural models where every atom has been assigned an explicit position.
% For disordered systems where preferred local structures are unknown, a common approach is to select structural models by applying some simple, and cheap to compute, heuristic; for example, ranking structures according to the electrostatic energies for a set of structurally equivalent point-charge models \cite{OngEtAl_EnergyEnvironSci2013, LuAndCiucci_JMaterChemA2018, LeeEtAl_RSCAdv2022}.
% Understanding classes of materials where more complex physics can direct local structure helps to define the limits of simpler structure-searching approaches, and may allow the development of better approximate structure-ranking schemes.

The antiperovskite-structured lithium oxychalcogenides \ce{(Li2$M$)\!\emph{Ch}O} (\!\emph{M} = a transition metal; \emph{Ch} = S, Se) are one family of cation-disordered materials that have been proposed as high-capacity cathodes for lithium-ion batteries \cite{LaiEtAl_JAmChemSoc2017, MikhailovaEtAl_ACSApplEnergyMater2018, LuAndCiucci_JMaterChemA2018,LaiEtAl_InorgChem2018,GorbunovEtAl_InorgChem2020}.
The most promising of these is the oxysulfide \ce{Li2FeSO} \cite{LuAndCiucci_JMaterChemA2018,LaiEtAl_JAmChemSoc2017,MikhailovaEtAl_ACSApplEnergyMater2018}, which has a first-cycle capacity of \SI{275}{\milli\ampere\hour\per\gram} at C/10.
The practical use of \ce{Li2FeSO} as a cathode material is limited by capacity-fade on cycling, associated with progressive amorphisation \cite{GorbunovEtAl_InorgChem2020}.
The underlying mechanism of this cycling-induced amorphisation remains unclear; in part, because the atomic structure of pristine \ce{Li2FeSO} is not yet fully characterised.

\begin{figure}[tb]
  \centering
  \resizebox{6.5cm}{!}{\includegraphics*{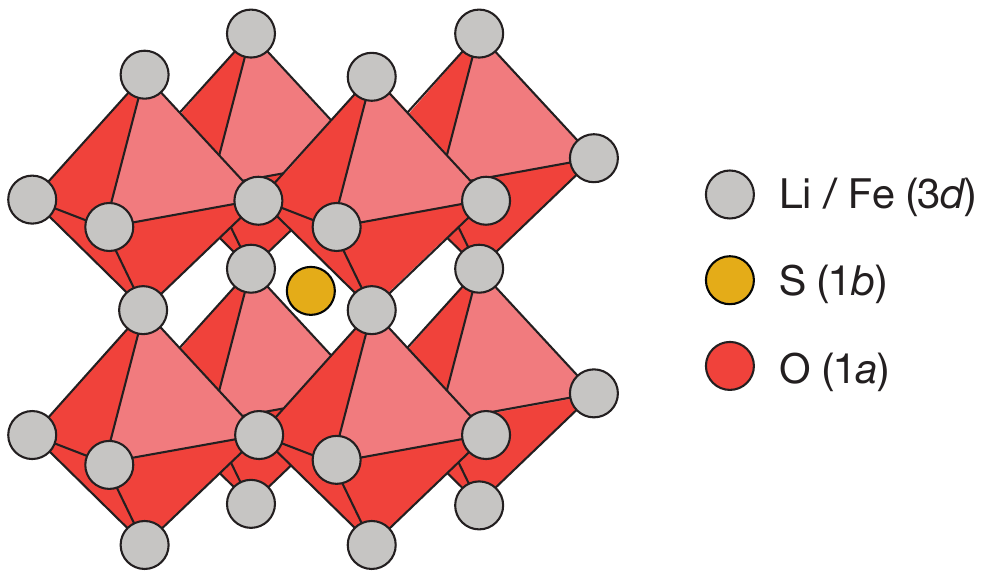}} %
    \caption{\label{fig:Li2FeSO_structure}The reported antiperovskite structure (space group $Pm\bar{3}m$) of \ce{Li2FeSO} \cite{LaiEtAl_JAmChemSoc2017}.
    Oxygen occupies the 6-coordinate octahedral Wyckoff $1a$ site. Sulfur occupies the 12-coordinate Wyckoff $1a$ site. Lithium and iron are distributed over the Wyckoff $3d$ sites in a 2:1 ratio.}
\end{figure}

Cubic antiperovskites, such as \ce{Li2FeSO}, are charge-inverted structural analogues of the well-known conventional cubic perovskites.
In \ce{Li2FeSO} (space group $Pm\bar{3}m$; $a=\SI{3.914}{\angstrom}$ \cite{LaiEtAl_JAmChemSoc2017}), sulfur occupies the 12-coordinate Wyckoff $1b$ site, oxygen occupies the 6-coordinate Wyckoff $1a$ site, and lithium and iron occupy the Wyckoff $3d$ sites in a 2:1 Li:Fe ratio (Fig.~\ref{fig:Li2FeSO_structure}).
Previous diffraction studies of \ce{Li2FeSO} have shown an absence of cation ordering at long ranges \cite{LaiEtAl_JAmChemSoc2017, MikhailovaEtAl_ACSApplEnergyMater2018}, leading to the assignment of Li and Fe as randomly distributed over the available Wyckoff $3d$ sites \cite{LaiEtAl_JAmChemSoc2017}.
%A fully random Li/Fe distribution requires that all Li/Fe configurations have equal energies.
While a ``fully random'' cation-distribution is consistent with the previous experimental diffraction data \cite{LaiEtAl_JAmChemSoc2017,MikhailovaEtAl_ACSApplEnergyMater2018}, alternative structural models with some preferential short-range order but no long-range order would also be compatible.
For mixed Li/Fe systems, such as \ce{Li2FeSO}, some degree of short-range order might in fact be expected:
Fe$^{2+}$ ions have a higher formal charge than Li$^{+}$ ions, and simple electrostatic arguments predict that cation configurations that maximise Fe--Fe separations should be energetically favoured, resulting in some form of short-range, and possibly even long-range, cation ordering.

While previous experimental data have been interpreted as evidence of random Li/Fe distribution in \ce{Li2FeSO} \cite{LaiEtAl_JAmChemSoc2017}, computational studies of \ce{Li2FeSO} have predicted that different Li/Fe configurations give different structural energies \cite{LuAndCiucci_JMaterChemA2018, ZhuAndScanlon_ACSApplEnergyMater2022}, indicating a preference for some specific cation configurations.
This apparent inconsistency between proposed structural models raises the question of whether the Li/Fe cation distribution in \ce{Li2FeSO} does in fact demonstrate preferential short-range order, and, if so, what form this takes.
Secondly, if \ce{Li2FeSO} does indeed exhibit short-range cation order, what is the physical origin of this cationic ordering, and can this be explained by, for example, simple models of point-charge electrostatics \cite{LuAndCiucci_JMaterChemA2018}?

To characterise the short-range order in \ce{Li2FeSO}, we have performed density functional theory (DFT) calculations, cluster-expansion--based Monte Carlo sampling, and X-ray total scattering experiments and pair-distribution function (PDF) analysis.
Our computational model predicts partial short-range ordering, characterised by a strong preference for \ce{OLi4Fe2} oxygen-coordination, with a weaker preference for \cis-\ce{OLi4Fe2} over \trans-\ce{OLi4Fe2} oxygen-coordination.
To validate this structural model, we have compared simulated pair-distribution functions (PDFs) against our experimental PDF data.
Our DFT-derived computational model gives better agreement with the experimental data than models generated assuming either a random Li/Fe distribution or using the ground-state structure from a simple point-charge electrostatic model.

We find that Li/Fe configurations that give polar anion-coordination are stabilised relative to configurations that give non-polar-anion coordination, when compared to the energy ranking predicted by a simple point-charge electrostatic model.
The stabilisation of polar Li/Fe configurations can be understood as a consequence of anions with polar coordination being electronically polarised, which lowers the net electrostatic energy for these configurations.
We attribute this anion-polarisation--induced cationic short-range order in \ce{Li2FeSO} as a consequence of high anion polarisabilities combined with a capacity for highly polar local cation configurations, and we expect this effect to be generally applicable in cation-disordered materials where the cations have different formal charges.

Finally, we discuss the role of preferential short-range order on the presence or absence of long-range order, and characterise this in terms of the configurational density of states and the temperature dependence of short- and long-range order parameters.
Preferential \cis-\ce{OLi4Fe2} oxygen coordination, as predicted by our DFT calculations, means that neighbouring pairs of \ce{OLi4Fe2} octahedra are configurationally underconstrained, and can adopt various different relative orientations.
This produces long-range disorder, in agreement with experimental diffraction data, and is associated with a continuous density of states, and short- and long-range order parameters vary continuously at any non-zero temperature.
In contrast, preferential \trans-\ce{OLi4Fe2} oxygen coordination fully constrains the relative orientations of neighbouring \ce{OLi4Fe2} pairs in two dimensions, producing long-range order.
In this case, short- and long-range order parameters show strong ordering to relatively high temperatures, before undergoing a more sudden change to partial disorder, characteristic of a formal order--disorder transition.

Beyond the specific case of \ce{Li2FeSO}, our results demonstrate how going beyond simple point-charge models can be necessary to understand local structure in such cation-disordered materials, and highlight the role of anion polarisation in directing short-range order, and consequently the presence or absence of long-range order, in these materials.

\begin{figure*}[tb]
  \centering
  \resizebox{18.0cm}{!}{\includegraphics*{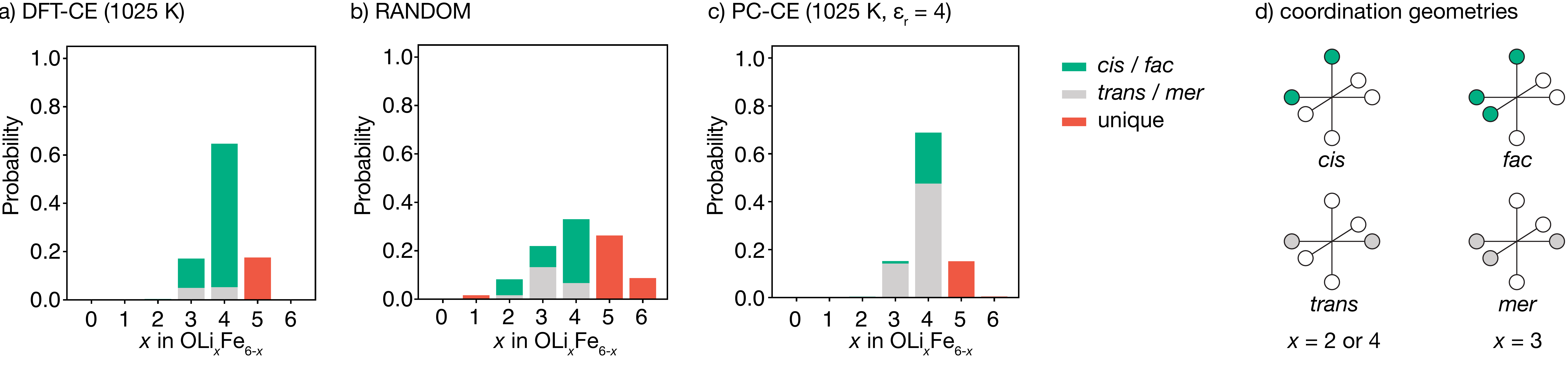}} %
    \caption{\label{fig:oxygen_coordination}(a--c) Oxygen coordination environment probability distributions for \ce{Li2FeSO} for (a) a density-functional theory derived cluster expansion model (DFT-CE), from Monte Carlo simulations at \SI{1025}{\kelvin}, (b) random Li/Fe positions ({RANDOM}), and (c) an on-site point-charge cluster expansion model (PC-CE), from Monte Carlo simulations at \SI{1025}{\kelvin}. (d) Schematic of \cis\ versus \trans\ (top) and \fac\ versus \mer\ (bottom) pseudo-octahedral coordination environments.}
\end{figure*}

\section{Results and Discussion}

\subsection{Quantifying short-range order in \ce{Li2FeSO}}
\label{sec:short-range-order}
The degree to which \ce{Li2FeSO} exhibits cationic short-range order can be quantified by considering the relative probabilities of different competing cation configurations.
Here, we consider the set of [Li$_x$Fe$_{6-x}$] oxygen coordination environments as a descriptor of possible short-range ordering.
To sample the different local configurations expected in as-synthesised \ce{Li2FeSO}, we first parametrise a cluster-expansion (CE) model from a set of DFT calculations with different \{Li,Fe\} configurations (see the Computational Methods section for details), and then use the resulting DFT-CE model to perform a Monte Carlo simulation in a $8\times8\times8$ supercell.
The relative probabilities of competing OLi$_x$Fe$_{6-x}$ coordination environments are obtained directly from analysis of the Monte Carlo simulation trajectory.

Our DFT-CE model (Fig.~\ref{fig:oxygen_coordination}a) predicts that the most likely oxygen coordination is \ce{OLi4Fe2}, which accounts for \SI{65}{\percent} of the oxygen environments.
Within this preferential \ce{OLi4Fe2} coordination, \SI{81}{\percent} of oxygen coordination environments are \cis-\ce{OLi4Fe2} and \SI{19}{\percent} are \trans-\ce{OLi4Fe2}.
We also predict moderate amounts of \ce{OLi3Fe3} and \ce{OLi5Fe2} oxygen-coordination, with the \ce{OLi3Fe3} environments divided 71:29 into \fac\ and \mer\ configurations.

For a fully random arrangement of Li and Fe, the corresponding oxygen-coordination probability distribution is a binomial distribution, with $p=\frac{2}{3}$ and $n=6$ (Fig.~\ref{fig:oxygen_coordination}b).
This distribution is visually distinct from that obtained from our Monte Carlo simulations, demonstrating the existence of short-range order in the DFT-parametrised computational model.

The preferential \ce{OLi4Fe2} coordination predicted by Monte Carlo simulation is consistent with the predictions of simple point-charge electrostatics.
Coordination environments that give ``local electroneutrality'' are generally expected to be favoured, as expressed by Pauling's second rule \cite{Pauling_JAmChemSoc1929}.
Assuming formal oxidation states, for corner-sharing O$^{2-}$[Li$^+_x$Fe$^{2+}_{6-x}$] octahedra local electroneutrality is achieved when $x=4$, corresponding to the preferential \ce{OLi4Fe2} coordination predicted by our DFT-CE model.

While a simple point-charge electrostatic model is consistent with preferential \ce{OLi4Fe2} coordination, this model also predicts \trans-\ce{OLi4Fe2} being favoured over \cis-\ce{OLi4Fe2}.
Considering Fe and Li as point-charges with formal 2+ and 1+ charges, respectively, that occupy ideal Wyckoff $3d$ crystallographic sites, the electrostatic energy of a \ce{OLi4Fe2} octahedron is minimised when the Fe ions maximise their separation, by occupying opposing octahedral vertices in a \trans\ configuration.
Indeed, the ground state for an on-site point-charge model is comprised of \SI{100}{\percent} \trans-\ce{OLi4Fe2} coordination.
At non-zero temperatures, entropic contributions mean some proportion of non-\trans-\ce{OLi4Fe2} oxygen coordination is expected.
Yet even at relatively high temperatures, a simple point-charge model  predicts a strong preference for \trans-\ce{OLi4Fe2} over \cis-\ce{OLi4Fe2} oxygen coordination---for a formal-charge on-site point-charge model for \ce{Li2FeSO} with relative permittivity $\eps=4.78$ at $T=\SI{1025}{K}$ (Fig.~\ref{fig:oxygen_coordination}c), the predicted \trans:\cis\ \ce{OLi4Fe2} ratio is 69:31.
Even in the high-temperature limit, where the point-charge model recovers the fully random model distribution, the ratio of \ce{OLi4Fe2} \trans\ versus \cis\ environments is 25:75, and therefore never reaches the ratio of 19:81 \trans- to \cis-\ce{OLi4Fe2} predicted by the DFT-CE model at \SI{1025}{K}.

\begin{figure}[htb]
  \centering
  \resizebox{7.9cm}{!}{\includegraphics*{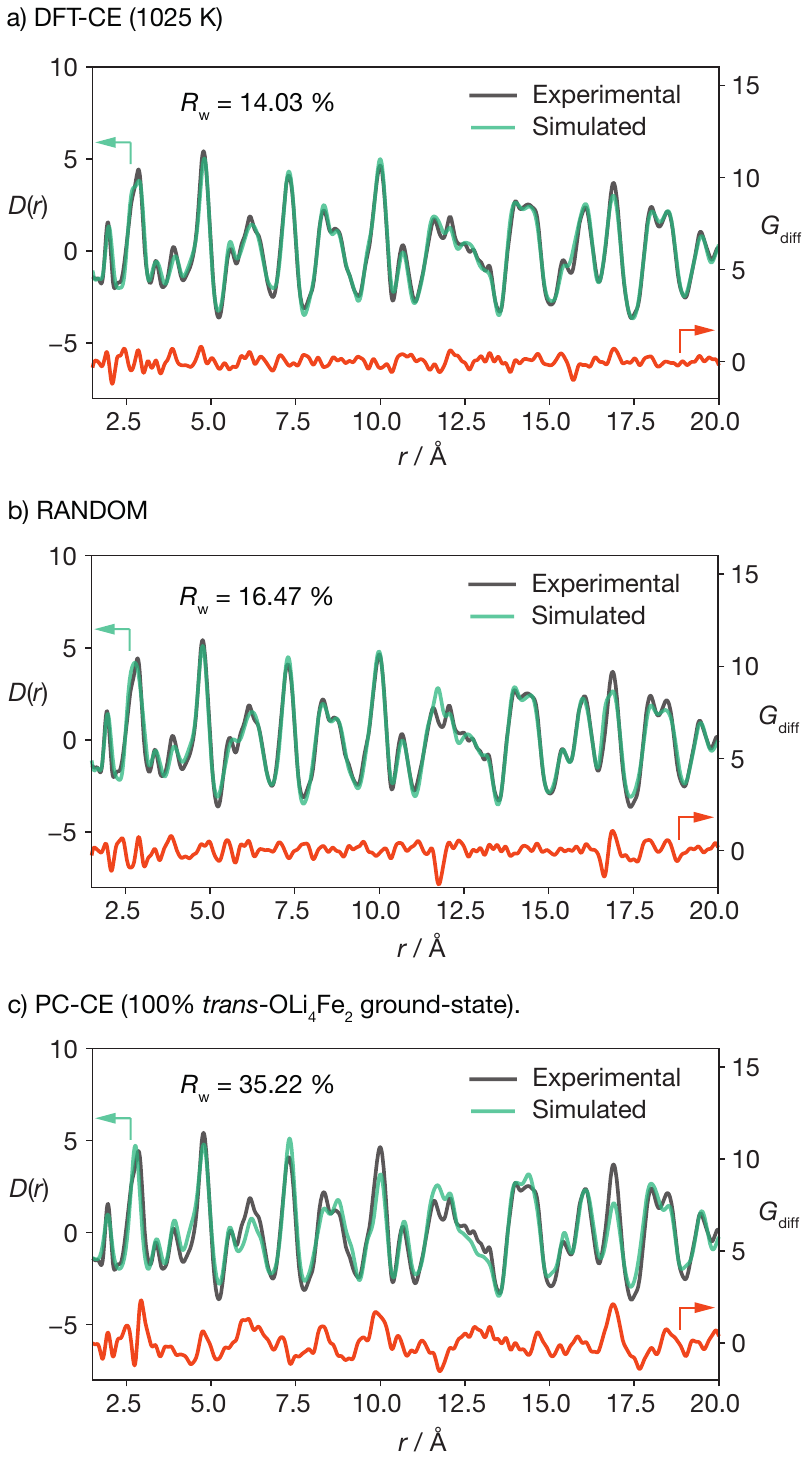}} %
    \caption{\label{fig:pdf_comparison}Comparison of pair-distribution functions (PDFs) obtained from X-ray total scattering and simulated PDF data for \ce{Li2FeSO} for structural models generated for (a) the  DFT-fitted cluster-expansion model (DFT-CE), (b) the RANDOM model, and (c) the PC-CE point-charge model.}
\end{figure}

To validate our DFT-CE model, we compared the pair distribution function (PDF) obtained from X-ray total scattering of \ce{Li2FeSO} to simulated PDFs for (a) the DFT-CE model, (b) the RANDOM model, and (c) the \SI{100}{\percent} \trans-\ce{OLi4Fe2} structure that would be obtained from a simple point-charge electrostatic ranking of all possible Li/Fe configurations (Fig.~\ref{fig:pdf_comparison}).
The best quality-of-fit is obtained for the DFT-CE model ($R_\mathrm{w}=\SI{14.03}{\percent}$), versus $R_\mathrm{w}=\SI{16.47}{\percent}$ for the RANDOM model, and $R_\mathrm{w}=\SI{35.22}{\percent}$ for the ground-state point-charge model.
The particularly poor quality-of-fit for the \SI{100}{\percent} \trans-\ce{OLi4Fe2} point-charge ground-state illustrates how a simple ranking of structures based on point-charge electrostatic energies would predict a structure that is incompatible with the experimental PDF data at short range ($<\SI{20}{\angstrom}$).

The oxygen-coordination populations obtained from the DFT-derived cluster expansion model indicate that, on average, structures with \cis-\ce{OLi4Fe2} are lower in energy than those with \trans-\ce{OLi4Fe2} coordination.
To better characterise the relative energies of competing \ce{OLi4Fe2} cation configurations as a function of the proportion of \cis- versus \trans-\ce{OLi4Fe2} oxygen coordination, we calculated the energies of all symmetry-inequivalent $2\times2\times2$ supercells containing only \ce{Li4Fe2} oxygen-coordination---i.e., every oxygen is coordinated by \ce{Li4Fe2} in either a \cis\ or \trans\ configuration.

\begin{figure}[tb]
  \centering
  \resizebox{8cm}{!}{\includegraphics*{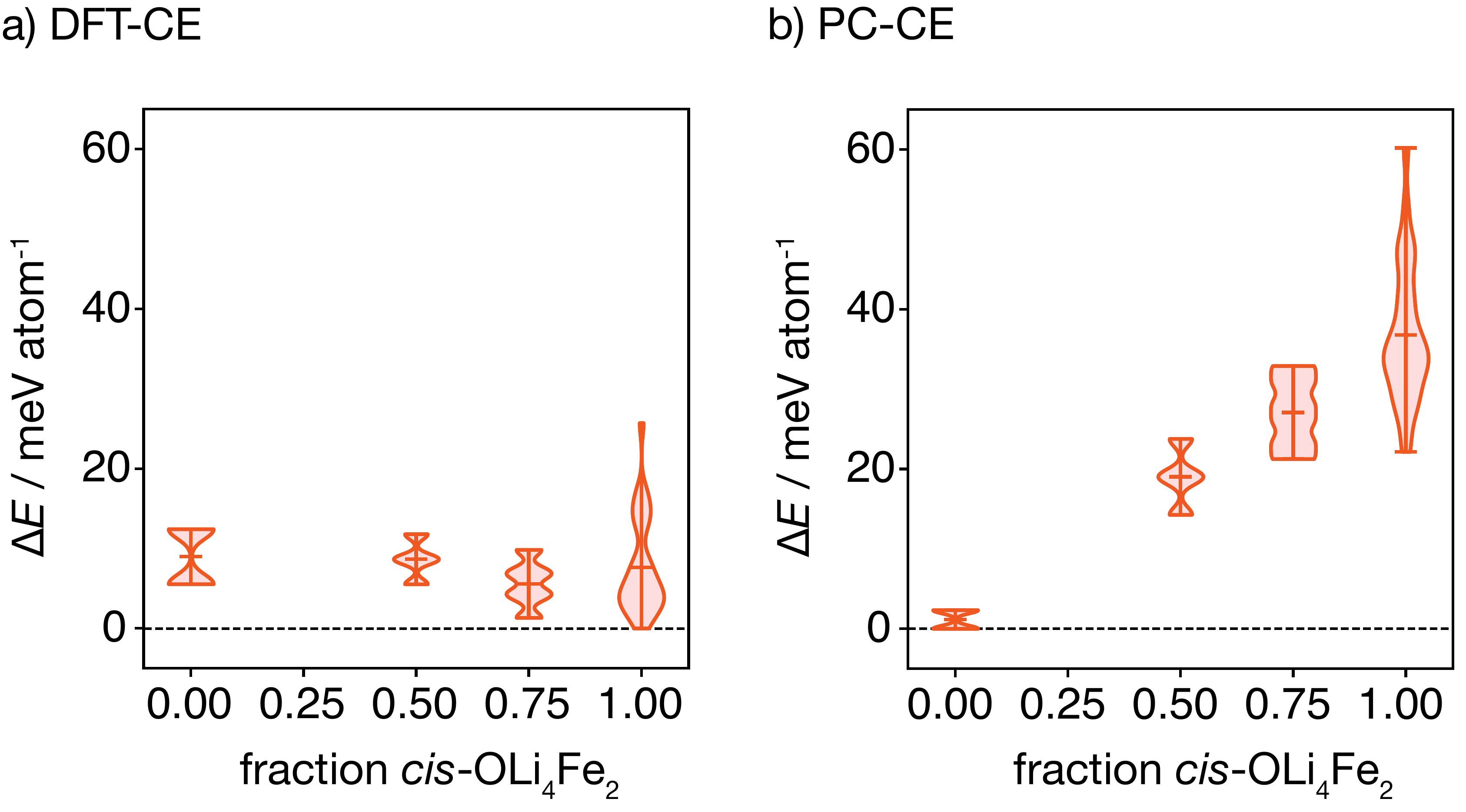}} %
    \caption{\label{fig:Li4Fe2_energies}Distributions of structural energies for all 2$\times$2$\times$2 structures consisting solely of \ce{OLi_4Fe_2} oxygen coordination environments, calculated using (a) the DFT-CE DFT-fitted cluster expansion model, and (b) the PC-CE point-charge model.
    For each model, the distributions are subdivided according to the fraction of \cis- versus \trans-\ce{OLi_4Fe_2}\ octahedra in each structure.}
\end{figure}

Fig.~\ref{fig:Li4Fe2_energies}(a) shows the distribution of these energies, relative to the ground-state, grouped by the proportion of \cis\ oxygen coordination environments in each structure.
The lowest energy structure has \SI{100}{\percent} \cis-\ce{OLi4Fe2} coordination, as expected from the oxygen-coordination probabilities obtained from Monte Carlo simulation.
Interestingly, the \emph{highest} energy structure containing only \ce{OLi4Fe2} coordination also has \SI{100}{\percent} \cis-oxygen coordination.
This suggests the existence of a more complex relationship between cation configuration and energy than a simple preference for \cis- versus \trans-oxygen-coordination.

Fig.~\ref{fig:Li4Fe2_energies}(b) shows the equivalent distribution of energies for all \SI{100}{\percent} \ce{OLi4Fe2} $2\times2\times2$ supercells for the point-charge model ($\eps=4.78$). Here, the lowest energy structures are \SI{100}{\percent} \trans-\ce{OLi4Fe2}, and the energy generally increases with increasing proportion of \cis-\ce{OLi4Fe2} coordination, as expected.

\subsection{Dipolar stabilisation of polar anion coordination}

\begin{figure}[tb]
  \centering
  \resizebox{8.5cm}{!}{\includegraphics*{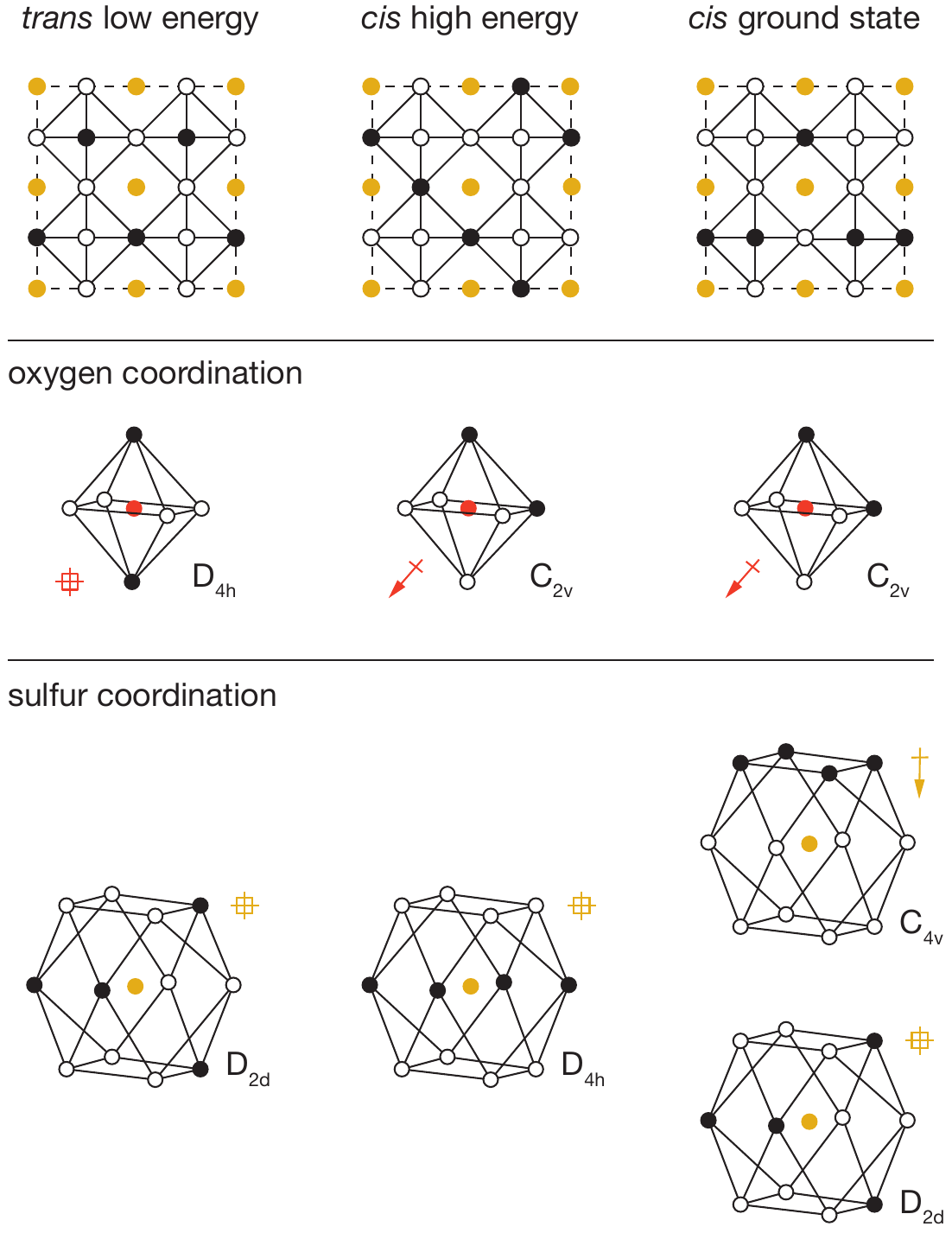}} %
    \caption{\label{fig:example_structures}Illustration of the three structures of interest (top) with their oxygen (middle) and sulfur (bottom) environments shown with them. For the illustrated coordination environments, polar coordination environments are marked with an arrow and non-polar enviroments are maked with a \nonpolar\ glyph.}
\end{figure}

To better understand the physical origin of the energy variation between structures with different \ce{OLi4Fe2} configurations, we performed further DFT calculations on three exemplar \ce{OLi4Fe2} $2\times2\times2$ configurations: the lowest energy all-\trans\ structure, the highest energy all-\cis\ structure, and the lowest energy all-\cis\ structure. 
These three structures are shown schematically in Fig.~\ref{fig:example_structures}, with their corresponding cation configurations around oxygen and sulfur.

In our analysis, we consider two possible contributions to the total energy that are not accounted for in a simple point-charge electrostatic model.
First, anions with polar coordination can move off their formal positions to give some shorter anion--cation distances, potentially resulting in stronger anion--cation interactions.
Second, anions with polar coordination will be electronically polarised, and these anion dipoles will lower the net electrostatic energy of the system relative to a simple sum over point-charge Coulomb terms.
This polarisation-stabilisation effect is potentially more significant in heterocationic materials, such as \ce{Li2FeSO}, where the central atom is a relatively more polarisable anion, than in conventional heteroanionic materials, such as \ce{NbO2F} \cite{BrinkEtAl_JSolStatChem2002} and \ce{SrNbO2N} \cite{YangEtAl_NatureChem2010}, where the central atom is typically a relatively unpolarisable high-formal-charge cation.

For each DFT-optimised structure, we consider the Fe--O and Fe--S distances and the magnitudes of the O and S electronic dipoles (Table \ref{table:structure_comparison}), which we calculate from Wannier analysis of the converged DFT calculations.
In the \trans\ structure, oxygen and sulfur both have non-polar coordination, and corresponding dipole moments of \SI{0.00}{\electron\angstrom}.
This structure can therefore be considered a reference system in which ionic and electronic polarisation effects are absent.
We also obtain reference O--Fe and S--Fe nearest-neighbour distances of \SI{1.97}{\angstrom} and \SI{2.74}{\angstrom} respectively.

In the \cis\ high-energy structure, the oxygen ions have polar coordination and exhibit a dipole moment of $\mu_\mathrm{O} =\SI{0.41}{\electron\angstrom}$. The sulfur ions have non-polar coordination, and a dipole moment of $\mu_\mathrm{S}=\SI{0.00}{\electron\angstrom}$.
The O--Fe nearest-neighbour distance is effectively unchanged from the value for the reference \trans\ structure, which suggests that the relative stability of this structure relative to the \SI{100}{\percent} \trans\ structure cannot be attributed to a change in O--Fe distance, and instead is due to the polarisation of oxygen anions in this \cis-\ce{OLi4Fe2}-coordination structure. 

In the \cis\ low-energy structure, all the oxygen ions and half the sulfur ions have polar coordination, and the calculated dipole moments for these polar-coordinated anions are $\mu_\mathrm{O}=\SI{0.33}{\electron\angstrom}$ and $\mu_\mathrm{S}=\SI{0.36}{\electron\angstrom}$, respectively. 
The polarisation of both oxygen and sulfur correlates with the lower energy of this \SI{100}{\percent} \cis-\ce{OLi4Fe2} structure relative to the \cis\ high-energy structure.
In the \cis\ high-energy structure, only the oxygen anions are polarised.
In the \cis\ ground-state structure, half the sulfur anions are also polarised, which provides additional dipole stabilisation, bringing the energy of this structure \emph{lower} than that of the \trans\ reference.
For the \cis\ ground-state structure, the O--Fe nearest-neighbour distance again is effectively unchanged from the reference \trans\ value, while the S--Fe nearest-neighbour distance decreases to \SI{2.68}{\angstrom}.
This off-site sulfur displacement potentially further stabilises this structure relative to the reference on-site point-charge model.

In heteroanionic oxyfluorides and oxynitrides, transition metals with polar six-coordinate \cis\ or \fac\ coordination move off-site to give shorter transition metal--oxygen distances, which stabilises these coordination geometries due to increased transition metal--oxygen covalent bonding \cite{KimEtAl_ChemMater2004,WithersEtAl_Polyhedron2007,YangEtAl_NatureChem2010,Attfield_CrysGrowDes2013,HaradaEtAl_AdvMater2019,PorterEtAl_JSolStatChem2015,WolffAndDronskowski_JComputChem2008}.
To characterise the degree of covalency in \ce{Li2FeSO}, and whether this varies with different short-range anion--cation coordination configurations, we calculated integrated crystal orbital bond index (ICOBI) \cite{MullerEtAl_JPhysChemC2021} values for all adjacent Fe--O pairs in each of our exemplar structures.
We obtain values of $\numrange{0.22}{0.25}$ for the three structures, which are indicative of \ce{Li2FeSO} being highly ionic \footnote{The ICOBI values reported here are comparable to those of ionic salts, such as \ce{LiCl} \cite{MullerEtAl_JPhysChemC2021}.} (see the Supporting Information for details).

\begin{table}[]
\begin{tabular}{llccc} % \toprule
\multicolumn{1}{l}{Structure}          & \multicolumn{1}{c}{$X$} & $r$(Fe--$X$) / \si{\angstrom} & $\left|\mu_X\right|$ / \si{\electron\angstrom} \\ \midrule
\multirow{2}{*}{\trans\ low-energy} & O$_\mathrm{non-polar}$ & 1.97 & 0.00 \\
                                    & S$_\mathrm{non-polar}$ & 2.74 & 0.00 \\ \cmidrule{2-4}
\multirow{2}{*}{\cis\ high-energy}  & O$_\mathrm{polar}$     & 1.96 & 0.41 \\
                                    & S$_\mathrm{non-polar}$ & 2.75 & 0.00 \\ \cmidrule{2-4}
\multirow{3}{*}{\cis\ ground-state} & O$_\mathrm{polar}$     & 1.99 & 0.33 \\
                                    & S$_\mathrm{polar}$     & 2.68 & 0.36 \\
                                    & S$_\mathrm{non-polar}$ & 2.73 & 0.00 \\
\end{tabular}\caption{Iron anion bond lengths and anion dipoles for the different types of anions in the different structures of interest.}\label{table:structure_comparison}
\end{table}

The observation that anions with polar coordination are themselves electronically polarised explains the average increased stability of \cis-\ce{OLi4Fe2} structures relative to \trans-\ce{OLi4Fe2} structures, when comparing DFT-predicted energies to the corresponding energies from the simple-point charge model (Fig.~\ref{fig:Li4Fe2_energies}).
The formation of dipoles lowers the energy of a given structure relative to the corresponding point-charge model energy, which does not account for electronic polarisation.
The \cis-\ce{OLi4Fe2} structures have polar oxygen coordination and exhibit strong oxygen polarisation, while the \trans-\ce{OLi4Fe2} structures have non-polar oxygen coordination and corresponding negligible oxygen polarisation.
On average, therefore, all \cis-\ce{OLi4Fe2} structures are more stable, relative to the \trans-\ce{OLi4Fe2} structures, than would be expected from simple point-charge electrostatics.

In structures where oxygen and sulfur both have polar coordination, both anions are polarised, giving greater net stabilisation relative to the point-charge model.
The effect of joint oxygen and sulfur polarisation is highlighted by comparing the relative energies of the \cis\ ground-state and \cis\ high-energy structures described above, predicted by the point-charge model versus the full DFT calculations.
The point-charge model predicts that the \cis\ ground-state structure has a \emph{higher} energy than the \cis\ high-energy structure ($\Delta E = \SI{36.0}{\meV\per\atom}$), while the DFT calculations predict that the \cis\ ground-state structure is more stable by $\Delta E = \SI{-25.7}{\meV\per\atom}$; we attribute this inversion of relative energies between the point-charge model and DFT data to the additional stabilising effect of sulfur polarisation in the \cis\ ground-state structure.

\subsection{Short-range ordering dictates long-range ordering}

The results presented above predict that \ce{Li2FeSO} exhibits short-range order characterised by a dual preference for \ce{OLi4Fe2} oxygen coordination and for polar cation coordination around the O and S anions, e.g., \cis-\ce{OLi4Fe2} over \trans-\ce{OLi4Fe2}.
The preference for polar cation-coordination around the anions is attributed to the capacity for these asymmetrically-coordinated anions to then polarise, with the resulting dipoles lowering the net electrostatic energy for these cation configurations.

\begin{figure}[tb!]
  \centering
  \resizebox{8.0cm}{!}{\includegraphics*{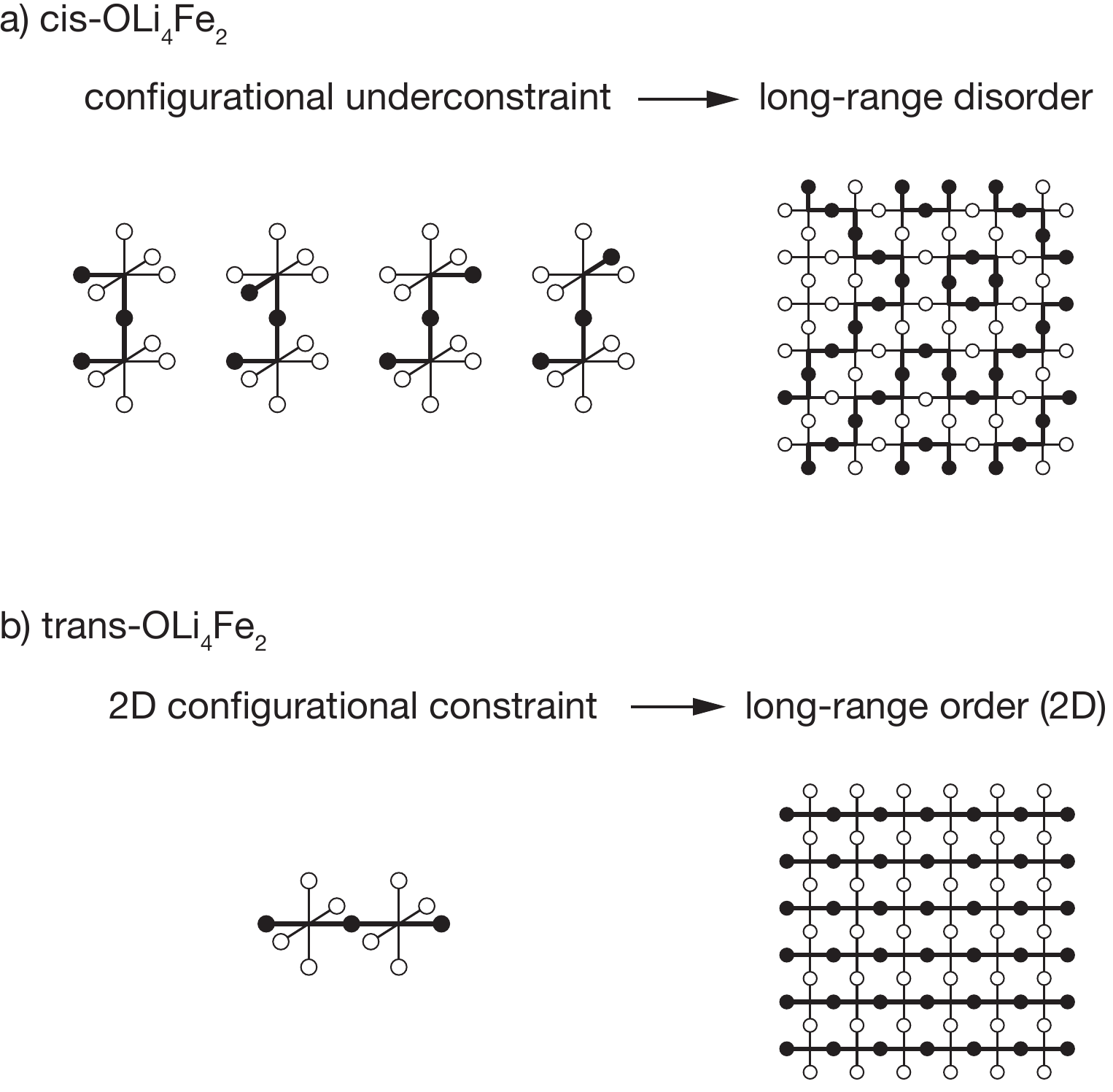}} %
    \caption{\label{fig:trans_vs_cis_ordering_schematic}Schematic illustrating how \cis\ versus \trans-\ce{OLi4Fe2} short-range order produces long-range disorder---in the case of an all-\cis\ ground-state---or long-range order---in the case of an all-\trans\ ground state---in the limit of $T=\SI{0}{\kelvin}$.}
\end{figure}

These short-range order preferences have a direct effect on the form and degree of long-range order in \ce{Li2FeSO}.
The role of preferential short-range order in directing long-range order is illustrated schematically in Fig.~\ref{fig:trans_vs_cis_ordering_schematic}. This figure shows examples for idealised \SI{100}{\percent} \cis\ versus \SI{100}{\percent} \trans-\ce{OLi4Fe2} coordination, i.e., the preferred local ordering as $T\to\SI{0}{K}$ for \ce{OLi4Fe2} as modelled using DFT and as predicted from a simple point-charge model, respectively.

For a Li/Fe configuration with \SI{100}{\percent} \cis-\ce{OLi4Fe2} oxygen coordination, the relative cation configurations around pairs of corner-sharing \ce{OLi4Fe2} octahedra are only partially correlated.
For any individual \cis-\ce{OLi4Fe2} octahedron, the neighbouring octahedra can adopt any of four possible configurations that still preserve the local \cis\ coordination.
This configurational underconstraint \cite{KeenAndGoodwin_Nature2015} means that even in a \SI{100}{\percent} \cis-\ce{OLi4Fe2} system, cation site occupations will progressively decorrelate with increasing cation--cation separation.
At long range the Fe and Li site occupations are fully uncorrelated  (Fig.~\ref{fig:trans_vs_cis_ordering_schematic}(a)), mirroring the lack of long-range cation order observed in diffraction experiments \cite{LaiEtAl_JAmChemSoc2017, MikhailovaEtAl_ACSApplEnergyMater2018}.

The configurational underconstraint exhibited in a \SI{100}{\percent} \cis-\ce{OLi4Fe2} system can be contrasted with the character of a \SI{100}{\percent} \trans-\ce{OLi4Fe2} system.
For this different short-range order preference, the relative orientation of each \trans-\ce{OLi4Fe2} coordinated octahedron fully constrains the orientations of four ``in-plane'' neighbouring octahedra (Fig.~\ref{fig:trans_vs_cis_ordering_schematic}b)
This short-range configurational constraint enforces long-range cation order in two dimensions; adjacent 2D planes can have mutually parallel or perpendicular relative orientations~\footnote{The lowest energy \SI{100}{\percent} \trans-\ce{OLi4Fe2} configuration for both the DFT-CE and PC-CE models corresponds to perpendicular Fe--O--Fe chains in each alternating 2D plane, which maximises the distance between Fe––Fe pairs in these adjacent layers.}.
Hence, a structure comprised of corner-sharing \ce{A2B4}-octahedra with a short-range preference for \trans-coordination is expected to be long-range ordered, while an otherwise equivalent structure with a short-range preference for \cis-coordination is expected to be long-range disordered \footnote{This effect, where the geometry of structural building blocks dictates the degree and form of long-range order in different materials, has previously been discussed by Overy \emph{et al} in the context of generalised ice rules \cite{OveryEtAl_NatureCommun2016}. The all-\cis\ and all-\trans\ structures discussed here (Fig.~\ref{fig:trans_vs_cis_ordering_schematic}) correspond to the \ce{C4C} and \ce{C4T} procrystalline systems, respectively.}.

\begin{figure}[tb]
  \centering
  \resizebox{8.6cm}{!}{\includegraphics*{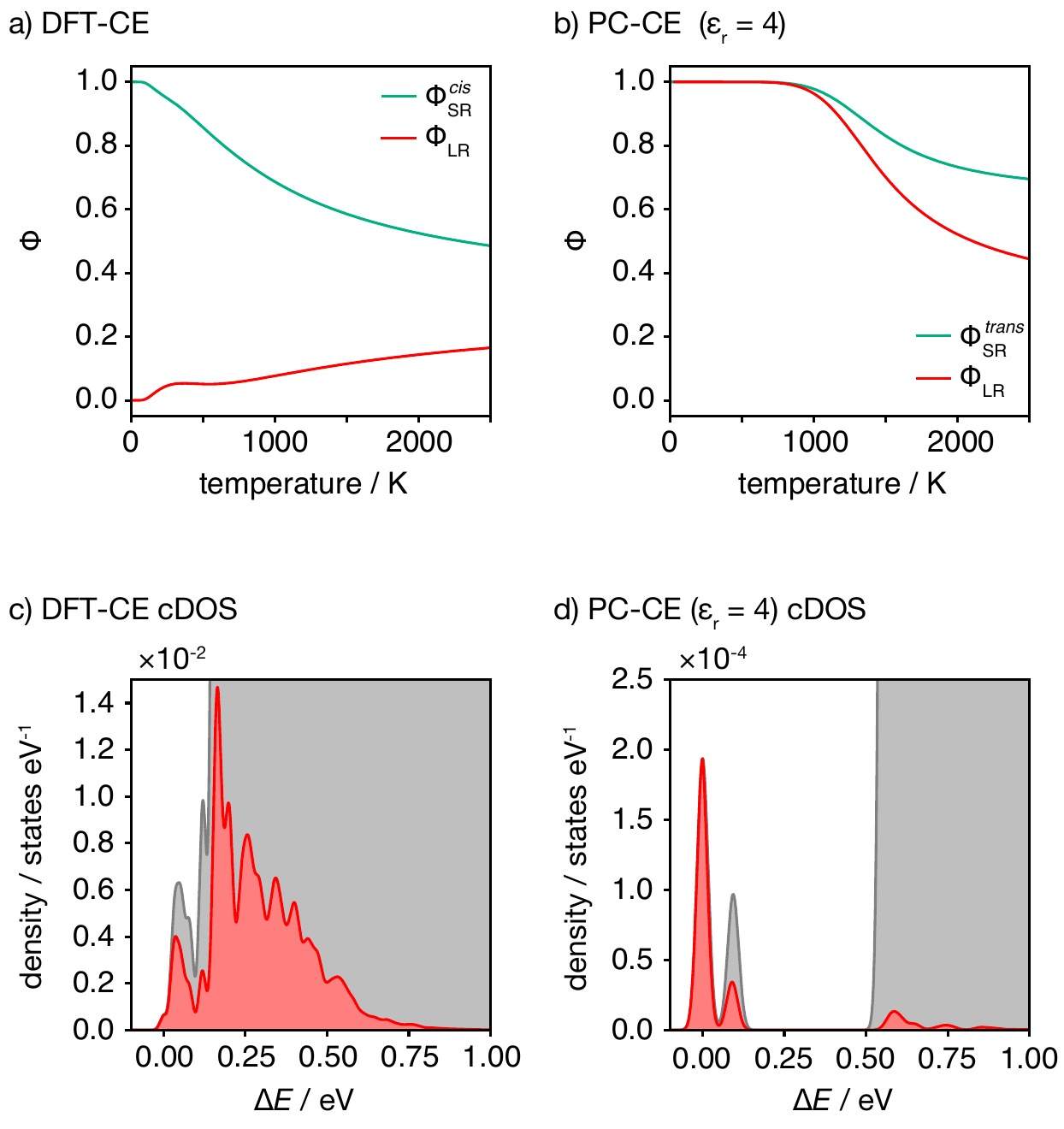}} %
    \caption{\label{fig:order_parameters_cdos}(a) and (b) Short- and long-range order parameters, $\Phi_\mathrm{SR}$ and $\Phi_\mathrm{LR}$ computed as a thermally weighted statistical average over all $2\times2\times2$ \ce{Li2FeSO} supercells for the DFT-CE and PC-CE ($\eps=4.78$) models. (c) and (d) Configurational densities of states (cDOS) for all $2\times2\times2$ \ce{Li2FeSO} supercells for the DFT-CE and PC-CE ($\eps=4.78$) models.
    Red shading indicates the distribution of thermally accessible states at $T=\SI{1025}{K}$, computed by Boltzmann-weighting the full cDOS.}
\end{figure}

This analysis of idealised \SI{100}{\percent} \cis- or \trans-\ce{OLi4Fe2} systems illustrates that in the limit of $T = \SI{0}{K}$ different short-range order preferences are predicted to lead to qualitative differences in the degree of long-range order.
In as-synthesised \ce{Li2FeSO} we expect a range of local coordination environments, as sampled in our Monte Carlo simulations.
The relationship between short-range order and long-range order as a function of configurational temperature can be quantified by considering appropriate short-range and long-range order parameters.
Fig.~\ref{fig:order_parameters_cdos}(a) and (b) show calculated short-range and long-range order parameters, $\Phi_\mathrm{SR}$ and $\Phi_\mathrm{LR}$, for the DFT-derived DFT-CE cluster expansion model and the on-site point-charge PC-CE model, respectively, calculated as a thermal average over all possible $2\times2\times2$ \ce{Li2FeSO} supercells.
For both models the long-range order parameter is defined as the proportion of collinear Fe--O--Fe units; in a $2\times2\times2$ supercell this is equivalent to long-range Fe--O--Fe--O--Fe ordering.
The short-range order parameter in each case is defined as the proportion of \cis\ or \trans\ \ce{OLi4Fe2} units in the DFT-CE and PC-CE models, respectively, i.e., it is the proportion of oxygen coordination environments that adopt the preferential $T=\SI{0}{K}$ short-range ordering.

For the DFT-CE model, at $T=\SI{0}{K}$, $\Phi_\mathrm{SR}=1$ and $\Phi_\mathrm{LR}=0$, as expected from the discussion above (Fig.~\ref{fig:trans_vs_cis_ordering_schematic}).
As the temperature is increased, the short-range order parameter starts to decrease, even at relatively low temperatures, and the long-range order parameter shows a small increase.
These short-range and long-range order parameters continuously decrease and increase, respectively, as the temperature is increased.
Even at relatively high temperature, however, the long-range order parameter $\Phi_\mathrm{LR}$ is still low, and there is no temperature regime where significant long-range ordering is predicted.
In contrast, for the PC-CE ($\epsilon_\mathrm{r}=4.78$) model, there is strong ordering at both short- and long-ranges until nearly $T=\SI{1000}{K}$.
Above $T\approx\SI{1000}{K}$ both order parameters start to decrease.
The correlation between $\Phi_\mathrm{SR}$ and $\Phi_\mathrm{LR}$ is expected because of the similarity in how these are defined for this system, and because of the long-range ordering promoted by preferential \trans-\ce{OLi4Fe2} coordination.

The DFT-CE and PC-CE models therefore show somewhat qualitatively different changes in their short-range order, and hence in their long-range order, as a function of temperature.
For the DFT-CE model, any increase in temperature above $T=\SI{0}{K}$ progressively disrupts the preferential \cis-\ce{OLi4Fe2} coordination.
For the PC-CE model, however, the short- and long-range order are resistant to thermal disordering up to $T=\SI{1000}{K}$.
This difference in behaviour can be understood by comparing the configurational densities of states (cDOS) for these two models (Fig.~\ref{fig:order_parameters_cdos}(c) and (d)).
The DFT-CE model gives a relatively narrow cDOS.
Not all cation configurations are thermally accessible at \SI{1025}{\kelvin}, but the distribution of thermally accessible states forms a continuous distribution that includes a relatively large number of inequivalent cation configurations.
For the point-charge PC-CE model, in contrast, the thermally accessible states at \SI{1025}{\kelvin} are largely restricted to a very narrow distribution at low energy that is split off from the other states by a large energy-gap.
This low energy peak consists entirely of \SI{100}{\percent} \trans-\ce{OLi4Fe2} configurations, and we therefore observe strong long-range ordering even at \SI{1025}{\kelvin}, with the sudden onset of disordering above this temperature indicative of a formal order--disorder transition.

\section{Summary and Discussion}

Here, we have reported a combined computational and experimental study of cation order in the cation-disordered heterocationic lithium-ion cathode material \ce{Li2FeSO}.
Our computational model predicts that lithium and iron cations in as-synthesised \ce{Li2FeSO} are not randomly distributed, as previously proposed based on experimental diffraction data \cite{LaiEtAl_JAmChemSoc2017}, but instead exhibit partial short-ranged order, characterised by a strong preference for \ce{OLi4Fe2} oxygen coordination environments, with a weaker preference for these to be \cis-\ce{OLi4Fe2} over \trans-\ce{OLi4Fe2}.
This result is supported by experimental X-ray pair-distribution function (PDF) analysis:
we show that our computational model gives better agreement with these experimental PDF data than either a fully-random cation distribution, or an ordered structure constructed on the basis of point-charge electrostatics.

To better understand the physical basis for these short-range order preferences, we performed an analysis of the energies of all possible $2\times2\times2$ \ce{Li2FeSO} supercells, using our DFT-derived cluster-expansion model, and compared these results to the corresponding energies predicted by a simple electrostatic model, wherein the Li and Fe cations are treated as point charges fixed at their formal crystallographic positions.
This analysis showed that, relative to the simple point-charge model, cation configurations with polar anion coordination, e.g., \cis-\ce{OLi4Fe2}, are stabilised relative to configurations with non-polar anion coordination, e.g., \trans-\ce{OLi4Fe2}.

% This stabilisation of Li/Fe configurations that give polar anion coordination can be understood as a consequence of the polarisation of these polar-coordination anions; the lowest energy cation configurations are those that allow both O and S polarisation.
% Anion polarisation therefore plays a critical role in directing short-range order in \ce{Li2FeSO}.
% Because the precise form of preferential short-range order directs the and, therefore, also explains the absence of long-range order in this material.

The stabilisation of polar anion coordination environments in heterocationic \ce{Li2FeSO} is functionally similar to the stabilisation of polar cation coordination environments in heteroanionic transition-metal oxyfluorides and oxynitrides \cite{HaradaEtAl_AdvMater2019}, such as \ce{NbO2F}\cite{BrinkEtAl_JSolStatChem2002} and \ce{SrNbO2N}\cite{YangEtAl_NatureChem2010} \ .
In \ce{Li2FeSO}, however, this stabilisation of lower-symmetry polar coordination environments appears to have a different physical origin.
In transition-metal oxyfluorides and oxynitrides, polar cation coordination environments are stabilised through off-centre displacements of the central transition-metal cation, to give enhanced covalent bonding between the central cation and the coordinating heteroanions \cite{WithersEtAl_Polyhedron2007,WolffAndDronskowski_JComputChem2008,YangEtAl_NatureChem2010,Attfield_CrysGrowDes2013}.
Here, we identify a different mechanism, whereby \emph{anions} with polar heterocationic coordination are electronically polarised, with the resulting dipoles lowering the net electrostatic energy of the system.
We expect this mechanism to be quite general for heterovalent (cations with different formal oxidation states) heterocationic materials.
This mechanism of anion-polarisation--mediated stabilisation of polar heterocationic coordination environments requires only that the central ion is polarisable, and that polar coordination configurations of the heterocations produce an electric field.
This provides a possible explanation for the difference in physical mechanism responsible for short-range order in \ce{Li2FeSO} versus previously studied heteroanionic transition-metal systems: the central ions in heterocationic materials are softer more-polarisable anions, while in heteroanionic materials the central ions are typically harder less-polarisable high-valence cations.

We have also examined how the precise form of preferential short-range order dictates the presence or absence of long-range order.
In \ce{Li2FeSO}, where the preferential short-range ordering gives configurationally underconstrained preferential \cis-\ce{OLi4Fe2} coordination, there is no long-range cation order, even at $T=\SI{0}{K}$.
We have contrasted this with the behaviour in a system where short-range ordering is directed purely by point-charge electrostatics.
In this case, the preferential short-range ordering gives configurationally constrained \trans-\ce{OLi4Fe2} coordination, and at $T=\SI{0}{K}$ the system is fully long-range ordered.
By comparing different models for \ce{Li2FeSO} we show that in the configurationally underconstrained DFT-predicted system, short- and long-range order parameters vary continuously at any non-zero temperature.
In contrast, in the configurationally constrained point-charge model, short- and long-range order parameters show strong ordering to relatively high temperatures, before the onset of partial disordering.
This temperature dependent onset of disorder is characteristic of a formal order--disorder transition, and can be understood as a consequence of a significant energy gap between fully ordered and disordered structures.

More generally, the results and analysis presented here illustrate how it can be necessary to go beyond simple point-charge models to predict or understand local structure in cation-disordered materials, and highlight the role of anion polarisation in directing short-range order and consequently explaining the presence or absence of long-range order.

\section{Methods}

\subsection{Computational Methods}
\label{sec:computational_methods}

All density functional theory (DFT) calculations were performed using the plane-wave DFT code VASP \cite{KresseAndJoubert_PhysRevB1999,KresseAndHafner_PhysRevB1994}.
Interactions between core and valence electrons were described using the projector-augmented-wave (PAW) method, \cite{Blochl_PhysRevB1994} with cores of [Mg] for Fe, [He] for O, [Ar] for S, and all electrons treated as valence for Li..
We used the GGA functional PBEsol with a Dudarev $+U$ correction applied to the Fe $d$ states (GGA$+U$), with $U_{\mathrm{Fe},d} =~$\SI{5.3}{eV} \cite{WangEtAl_PhysRevB2006}.
All calculations used a plane-wave basis-set cut-off of \SI{720}{eV}.
Reciprocal space was sampled using a minimum $k$-point spacing of \SI{0.25}{\per\angstrom}.
For each structure, the ionic positions and the cell parameters were relaxed until all atomic forces were less than \SI{1e-2}{\eV\per\angstrom}.
All calculations were spin-polarised and were initialised in ferromagnetic configurations, and then allowed to relax without electronic constraint.
All calculations remained ferromagnetic during the DFT geometry optimisation.

To quantify anion polarisation in select structures we performed additional post-processing to compute the set of maximally-localised Wannier functions \cite{MarzariEtAl_RevModPhys2012} using the \texttt{Wannier90} code \cite{PizziEtAl_JPhysCondensMatter2020}.
Dipoles on the ions of the cathode material are obtained by associating Wannier centres with ions and calculation of the dipole from the vectors between positively charged ionic cores and the negatively charged ionic centres.
Full details are provided in the supporting information [reference to be added at publication].

To allow the computationally efficient evaluation of relative energies of \ce{Li2FeSO} structures with different Li/Fe configurations, we parametrised a cluster-expansion effective Hamiltonian \cite{SanchezEtAl_PhysAStatMechAppl1984,Sanchez_PhysRevB2010} by fitting to the DFT-calculated energies for 111 \ce{Li2FeSO} configurations, using the \textsc{icet} and \textsc{trainstation} packages \cite{AngqvistEtAl_AdvTheorySimul2019}, with a limit of a maximum of 40 non-zero features.
We use the Least Absolute Shrinkage and Selection Operator (LASSO), in combination with recursive feature elimination.
The resulting cluster-expansion model gives a cross-validation score of \SI{8}{meV/atom}.
To construct our simple point-charge model, we fit a second cluster expansion Hamiltonian (PC-CE) to the energies obtained from an Ewald sum with a relative permittivity of $\eps=4.78$ \footnote{$\eps=4.78$ is the electronic contribution to the static dielectric constant, calculated using dielectric perturbation theory \cite{GajdosEtAl_PhysRevB2006} using the HSE06 hybrid functional and a $4\times4\times4$ k-point grid.}, for ions with formal charges positioned at their corresponding crystallographic sites.
This model was fitted against all symmetry inequivalent arrangements of ions within $2\times1\times1$ and $2\times2\times1$ supercells \cite{Morgan_JOpenSourSoft2017}.
Both cluster expansions were fit using the the standard sinusoidal basis function used by \textsc{icet}, and considered all possible two, three, and four-body terms within cutoffs of \SI{15}{\angstrom}, \SI{9}{\angstrom}, and \SI{5}{\angstrom} respectively.
For both cluster expansion models, this fitting procedure gave no non-zero four-body terms (a full description of the fitted cluster-expansion weights is available as part of the supporting data repository \cite{data_Li2FeSO_structure_github}).

To investigate the influence of magnetic ordering on the predictive accuracy of our cluster expansion model, we considered the three exemplar structures discussed in detail in the main manuscript and performed additional calculations imposing antiferromagnetic ordering.
For all three calculations, antiferromagnetic (AFM) ordering was predicted to be more stable than ferromagnetic (FM) ordering by between \SI{7}{\meV\per\atom} and \SI{10}{\meV\per\atom}, in agreement with the equivalent analysis reported in Ref.~\onlinecite{ZhuAndScanlon_ACSApplEnergyMater2022}.
For our cluster expansion model we are interested in relative energies of different Li/Fe configurations.
The change in relative energy for the three test structures produced by using energies with AFM ordering rather then FM ordering is $<\SI{2}{\meV\per\atom}$ (further details are given in the Supplementary Information), which is both much smaller than the cross-validation score for our CE model of \SI[per-mode=power]{8}{\meV\per\atom}, and is negligible at our Monte Carlo simulation temperature of \SI{1050}{\kelvin}.

To model the probable distribution of different Li/Fe configurations, we performed lattice Monte Carlo simulations using our parametrised cluster-expansion Hamiltonian, using the \texttt{mchammer} software package \cite{AngqvistEtAl_AdvTheorySimul2019}.
These Monte Carlo simulations were performed in the canonical ensemble for $8\times8\times8$ supercells using both the DFT and electrostatic fitted Hamiltonians.
Initial configurations were generated at random, and then annealed from \SI{20000}{\kelvin} to \SI{1025}{\kelvin}, which corresponds to the experimental synthesis temperature, with \num{500000} attempted steps (approximately \num{32.5} MC Cycles), followed by a production run at \SI{1025}{\kelvin} of \num{1000000} attempted steps (approximately \num{641} MC Cycles).
Our reference distributions for fully random Li/Fe configurations were generated from a sample of \num{1000} random arrangements of cations.
For the modelling of PDF data we generated $4\times4\times4$ supercells using this same procedure, that were then relaxed using DFT using the same protocol as for the initial training set.
For PDF modelling of a ``random'' Li/Fe distribution we used a special quasi-random structure \cite{vandeWalleEtAl_Calphad2013} within the same $4\times4\times4$ supercell.

The three structures used for comparison with the experimental PDF were generated as follows: the $4\times4\times4$ DFT-CE structure was generated from lattice Monte Carlo simulations using the \textsc{icet} library, following the same MC protocal as described above; the $4\times4\times4$ RANDOM structure was generated as a special quasi-random structure using \textsc{icet}; the ordered \trans-structure is the electrostatic ground state, as obtained from a full enumeration of all $2\times2\times2$ cells, with energies calculated by Ewald summation using \textsc{pymatgen}.

\subsection{Synthesis and X-ray total scattering methods}

For experimental characterisation, samples of \ce{Li2FeSO} were prepared from stoichiometric amounts of \ce{Li2O} (Alfa Aesar, \SI{99.5}{\percent}), Fe (Alfa Aesar, \SI{99.9}{\percent}), and S (Alfa Aesar, \SI{99.5}{\percent}) using a slightly modified version of the previously reported method \cite{LaiEtAl_JAmChemSoc2017}.
The homogenised reactants were pressed into pellets, transferred in alumina crucibles and sealed within fused silica tubes under vacuum.
Deviating from the previously reported procedure, the samples were placed directly in preheated furnaces at \SI{750}{\celsius}.
The pellets were annealed for \SI{4}{\hour} at this temperature, with intermittent grinding, and subsequently quenched in ice water.
The resulting product was manually ground to obtain a fine powder.
All handling of the starting materials and products was performed under dry inert gas atmosphere in an Ar-filled glove box. 

X-ray total scattering data were collected at beamline I15-1 at the Diamond Light Source with an X-ray beam of energy \SI{76.69}{\kilo\electronvolt} ($\lambda=\SI{0.1617}{\angstrom}$) and a PerkinElmer XRD 1611 CP3 area detector.
Data reduction and normalisation were performed using \textsc{DAWN} \cite{BashamEtAl_JSynchroRad2015} and GudrunX \cite{SoperAndBarney_JApplCryst2011, Soper:2012} respectively, with $Q_{\textrm{min}}=\SI{0.5}{\per\angstrom}$ and $Q_{\textrm{max}}=\SI{28.0}{\per\angstrom}$.
Pair distribution function (PDF) refinements were performed using the PDFgui software \cite{FarrowEtAl_JPhysCondensMatter2007}.
PDF fits were performed in the range $1.5\,\leq r \leq20\,$\AA\; the following parameters were refined in each case: scale factor, lattice parameters $a$, $b$, and $c$, atomic correlation factor, and isotropic displacement parameters for each element.
A \ce{LiFeO2} disordered rocksalt side phase was identified from conventional Rietveld analysis and was included in the real-space refinements.

\section{Data and code availability}
A dataset containing the inputs and outputs for all the DFT calculations described in this paper is available from the University of Bath Research Data Archive \cite{data_Li2FeSO_DFT}.
All code used for analysis of our raw DFT data is available on GitHub \cite{data_Li2FeSO_structure_github}. This analysis uses the \textsc{bsym} \cite{Morgan_JOpenSourSoft2017}, \textsc{polyhedral-analysis} \cite{morgan_bjmorganpolyhedral-analysis_2020}, \textsc{scipy} \cite{VirtanenEtAl_NatMethods2020}, \textsc{matplotlib} \cite{Hunter_ComputSciEng2007}, \textsc{numpy} \cite{HarrisEtAl_Nature2020}, \textsc{pymatgen} \cite{OngEtAl_CompMatSci2013}, and \textsc{ase} \cite{HjorthLarsenEtAl_JPhysCondensMatter2017} packages. 

\section{Acknowledgements}
The authors thank the Faraday Institution CATMAT and FutureCat projects (EP/S003053/1, FIRG016, FIRG017) for financial support.
This work used the Michael computing cluster (FIRG019), the ARCHER2 UK National Supercomputer Service, with access provided by our membership of the UK's HPC Materials Modelling Consortium (EP/R029431), and the Isambard 2 UK National Tier-2 HPC Service operated by GW4 and the UK Met Office, and funded by EPSRC (EP/T022078/1).
The authors acknowledge the Diamond Light Source for allocation of beamtime under proposal number CY27702.
BJM thanks the Royal Society for financial support (URF/R/191006).
AGS thanks the STFC Batteries Network for an Early Career Researcher Award (ST/R006873/1).
HSG and ALG acknowledge funding from the ERC (grant 788144).

\bibliography{new}% Produces the bibliography via BibTeX.

\end{document}

\sisetup{per-mode = reciprocal, bracket-unit-denominator = true, sticky-per}%

% \preprint{APS/123-QED}

\title{Supplementary Information: Anion-polarisation--directed short-range-order in antiperovskite Li$_2$FeSO}

\author{Samuel W. Coles}
\email{swc57@bath.ac.uk}
\affiliation{Department of Chemistry, University of Bath, Claverton Down, BA2 7AY, United Kingdom}
\affiliation{The Faraday Institution, Quad One, Harwell Science and Innovation Campus, Didcot, OX11 0RA, United Kingdom}

\author{Viktoria Falkowski}
\affiliation{Department of Chemistry, University of Oxford, Inorganic Chemistry Laboratory, Oxford OX1 3QR, United Kingdom}
\affiliation{The Faraday Institution, Quad One, Harwell Science and Innovation Campus, Didcot, OX11 0RA, United Kingdom}

\author{Harry S. Geddes}
\affiliation{Department of Chemistry, University of Oxford, Inorganic Chemistry Laboratory, Oxford OX1 3QR, United Kingdom}
\affiliation{The Faraday Institution, Quad One, Harwell Science and Innovation Campus, Didcot, OX11 0RA, United Kingdom}

\author{Gabriel E. P\'erez}
\affiliation{ISIS Neutron and Muon Source, STFC Rutherford Appleton Laboratory, Didcot OX11 0QX, United Kingdom}
\affiliation{The Faraday Institution, Quad One, Harwell Science and Innovation Campus, Didcot, OX11 0RA, United Kingdom}

\author{Samuel G. Booth}
\affiliation{Department of Materials Science and Engineering, University of Sheffield, Sheffield S1 3JD, United Kingdom}
\affiliation{The Faraday Institution, Quad One, Harwell Science and Innovation Campus, Didcot, OX11 0RA, United Kingdom}

\author{Alexander G. Squires}
\affiliation{Department of Chemistry, University of Bath, Claverton Down, BA2 7AY, United Kingdom}
\affiliation{Department of Chemistry, University College London, London WC1H 0AJ, United Kingdom}
\affiliation{The Faraday Institution, Quad One, Harwell Science and Innovation Campus, Didcot, OX11 0RA, United Kingdom}

\author{Conn O'Rourke}
\affiliation{Department of Chemistry, University of Bath, Claverton Down, BA2 7AY, United Kingdom}
\affiliation{The Faraday Institution, Quad One, Harwell Science and Innovation Campus, Didcot, OX11 0RA, United Kingdom}

\author{Kit McColl}
\affiliation{Department of Chemistry, University of Bath, Claverton Down, BA2 7AY, United Kingdom}
\affiliation{The Faraday Institution, Quad One, Harwell Science and Innovation Campus, Didcot, OX11 0RA, United Kingdom}

\author{Andrew L. Goodwin}
\affiliation{Department of Chemistry, University of Oxford, Inorganic Chemistry Laboratory, Oxford OX1 3QR, United Kingdom}
\affiliation{The Faraday Institution, Quad One, Harwell Science and Innovation Campus, Didcot, OX11 0RA, United Kingdom}

\author{Serena A. Cussen}
\affiliation{Department of Materials Science and Engineering, University of Sheffield, Sheffield S1 3JD, United Kingdom}
\affiliation{The Faraday Institution, Quad One, Harwell Science and Innovation Campus, Didcot, OX11 0RA, United Kingdom}

\author{Simon J. Clarke}
\affiliation{Department of Chemistry, University of Oxford, Inorganic Chemistry Laboratory, Oxford OX1 3QR, United Kingdom}
\affiliation{The Faraday Institution, Quad One, Harwell Science and Innovation Campus, Didcot, OX11 0RA, United Kingdom}

\author{M. Saiful Islam}
\affiliation{Department of Chemistry, University of Bath, Claverton Down, BA2 7AY, United Kingdom}
\affiliation{Department of Materials, University of Oxford, Oxford, OX1 3PH, United Kingdom}
\affiliation{The Faraday Institution, Quad One, Harwell Science and Innovation Campus, Didcot, OX11 0RA, United Kingdom}

\author{Benjamin J. Morgan}
\email{b.j.morgan@bath.ac.uk}
\affiliation{Department of Chemistry, University of Bath, Claverton Down, BA2 7AY, United Kingdom}
\affiliation{The Faraday Institution, Quad One, Harwell Science and Innovation Campus, Didcot, OX11 0RA, United Kingdom}

% \collaboration{MUSO Collaboration}%\noaffiliation

\date{\today}% It is always \today, today,

\maketitle

%\tableofcontents

\section{Integrated Crystal Orbital Bond Index (ICOBI) analysis}
In the main manuscript we report a preference for \cis-\ce{OLi2Fe4} over \trans-\ce{OLi2Fe4} oxygen coordination in \ce{Li2FeSO}.
Many heteroanionic materials exhibit a similar preference for \cis\ coordination of anions, which is often attributed to \cis\ coordination giving a stronger covalent interaction between anions and transition metal ions, due $\pi$-bonding between anion p and transition-metal d orbitals  \cite{WithersEtAl_Polyhedron2007,YangEtAl_NatureChem2010,Attfield_CrysGrowDes2013,HaradaEtAl_AdvMater2019,WolffAndDronskowski_JComputChem2008,LegeinEtAl_InPrep2022}.
Here, we characterise the degree of ``covalency'' in \ce{Li2FeSO} by calculating integrated crystal orbital bond index (ICOBI) values \cite{MullerEtAl_JPhysChemC2021} for the three exemplar structures described in the main manuscript (Table.~\ref{table:structure_comparison}).
Smaller ICOBI values indicate more ``ionic'' bonding.
The absolute ICOBI values are similar to those calculated for LiCl, indicating that \ce{Li2FeSO} is highly ionic \cite{MullerEtAl_JPhysChemC2021}.
By comparison, the Ti--O bonds in \ce{BaTiO3} have an ICOBI value 3 times higher than we obtain for \ce{Li2FeSO} \cite{MullerEtAl_JPhysChemC2021}.
This high ionic bond-character is consistent with our\textbf{} proposal that the preferential \cis-\ce{OLi4Fe2} short-range ordering in \ce{Li2FeSO} is directed by electrostatics---specifically, anion-polarisation of anions with polar coordination environments, with a resulting electrostatic stabilisation of these coordination motifs.

\begin{table}[h!]
\begin{tabular}{lccc} % \toprule
\multicolumn{1}{l}{Structure}      &    $r$(Fe--$X$) / \si{\angstrom} & ICOBI (Fe--O)   \\ \midrule
\multirow{1}{*}{\trans\ low-energy}  & 1.97 & 0.23 & \\
\multirow{1}{*}{\cis\ high-energy}   & 1.96 & 0.25 & \\
\multirow{1}{*}{\cis\ ground-state}  & 1.99 & 0.22 \\
\end{tabular}\caption{Iron--oxygen bond lengths and ICOBI values for the different structures of interest.}\label{table:structure_comparison}
\end{table}

\section{Projected densities of states}

\begin{figure}[h!]
  \centering
  \resizebox{8.5cm}{!}{\includegraphics*{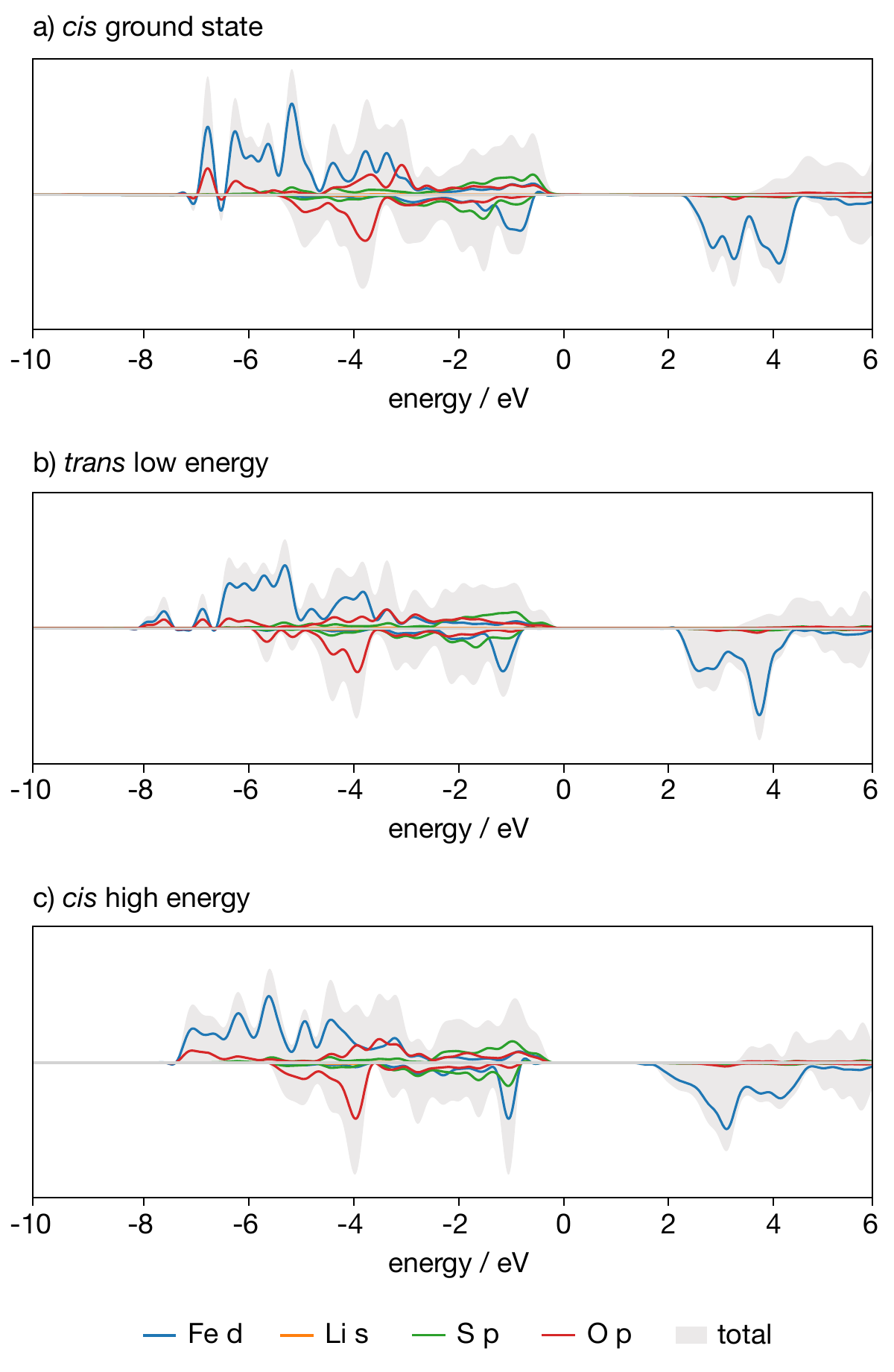}} %
    \caption{\label{fig:pdos}Projected densities of states (pDOS) of the \emph{cis} ground state structure (top panel), \emph{cis} high-energy structure (middle panel), and \emph{trans} low-energy structure (bottom panel).}
\end{figure}

Fig.~\ref{fig:pdos} shows projected densities of electronic states (pDOS) for the three exemplar \ce{Li2FeSO} structures analysed in the main manuscript.

\section{Energetic differences due to magnetic ordering}

Table.~\ref{table:AFM} lists the energy differences between ferromagnetic (FM) and antiferromagnetic (AFM) spin configurations for the three exemplar structures discussed in the main manuscript.

\begin{table}[h!]
\begin{tabular}{cc} % \toprule
Structure  &    $\Delta E_{\mathrm{FM}\to\mathrm{AFM}}$ / meV atom$^{-1}$  \\ \hline
\trans\ low-energy  & -7.2  \\
\cis\ high-energy   & -10.0  \\
\cis\ ground-state  & -8.4 \\
\end{tabular}\caption{Stabilisation energy for AFM versus FM spin ordering, $\Delta E_{\mathrm{FM}\to\mathrm{AFM}}$, for the three exemplar \ce{Li2FeSO} structures analysed in the main text.}\label{table:AFM}
\end{table}

\bibliography{new}% Produces the bibliography via BibTeX.